\documentclass[aps,prd,groupedaddress,showpacs,showkeys]{revtex4}
\usepackage{epsfig}
\newcommand{\eq}{\begin{eqnarray}}
\newcommand{\en}{\end{eqnarray}}
\newcommand{\ba}[1]{\begin{eqnarray} \label{(#1)}}
\newcommand{\ea}{\end{eqnarray}}

\def\p{\prime}

\def\rpm{R_p \hspace{-0.8em}/\;\:}
\def\rp{$R_p\hspace{-1em}/\ \ $}

\begin{document}

\title{On the possible resolution of the B-meson decay \\
       polarization anomaly in R-parity violating SUSY}
\author{
Amand Faessler$^1$,
Thomas Gutsche$^1$,
J.C. Helo$^2$, Sergey Kovalenko$^2$,
Valery E. Lyubovitskij$^1$
\footnote{On leave of absence
from Department of Physics, Tomsk State University,
634050 Tomsk, Russia}
\vspace*{1.2\baselineskip}}

\affiliation{\mbox{}\\
$^1$ Institut f\"ur Theoretische Physik,
Universit\"at T\"ubingen,
\\ Auf der Morgenstelle 14, D-72076 T\"ubingen, Germany
\vspace*{1.2\baselineskip} \\
\hspace*{-1cm}$^2$ Centro de Estudios Subat\'omicos(CES),
Universidad T\'ecnica Federico Santa Mar\'\i a, \\
Casilla 110-V, Valpara\'\i so, Chile
\vspace*{1.2\baselineskip} \\}

\date{\today}

\begin{abstract}

We examine the possible resolution of the recently observed
polarization anomaly in $B^0(\bar{B}^0)\rightarrow
\phi K^{*0}(\bar{K}^{*0})$-decay within R-parity
violating (\rp) SUSY.
We show that a combination of the superpotential trilinear
\rp-interactions, with the couplings $\lambda^{\prime}$, and
the soft SUSY breaking bilinear \rp \, sneutrino-Higgs mixing,
proportional to $\tilde{\mu}^2$, can potentially generate
the effective operators with the chirality structure necessary
to account for this anomaly.
However, we demonstrate that the existing experimental data on
$B_s\rightarrow \mu^+\mu^-$-decay lead to stringent upper limits on
the Wilson coefficients of these operators, which are about two orders
of magnitude below the values required for the resolution of the
B-decay polarization anomaly, and, therefore, it can hardly be
explained within the \rp SUSY framework.

As a byproduct result of our analysis we derive new limits on the
products of the soft bilinear and the superpotential trilinear
\rp-parameters of the form $\tilde{\mu}^2 \lambda^{\prime}$.

\end{abstract}



\maketitle

\newpage

\section{Introduction}

Now it is widely recognized that B-mesons offer powerful means for
testing the standard model (SM) and probe physics beyond its framework.
Recently, remarkable progress has been achieved in experimental and
theoretical studies of B-physics.
One of the most important experimental results of the last years in this
field was, certainly, the discovery of CP violation in the B-system.
The running B-experiments~\cite{PDG} at BABAR, BELLE, CDF, D0 and
CLEO have also collected a large statistics on various decay modes of
B-mesons some of which seem to be quite challenging for the SM.

The BABAR~\cite{BABAR} and BELLE~\cite{BELLE} Collaborations reported
experimental data on B-meson decay to a pair of light vector mesons:
$B\rightarrow VV$ where $V=\rho, \phi, K^*$.
An intriguingly large transverse polarization fraction comparable to
the longitudinal one has been observed in the
$B^0(\bar{B}^0)\rightarrow \phi K^{*0}(\bar{K}^{*0})$-decay channel.
This result has been recently confirmed by the CDF collaboration~\cite{CDF}
as well. This polarization anomaly is hard to be explained within the SM
and may indicate some new physics. As is known, the SM predicts
for the helicity amplitudes of $B^0\rightarrow \phi K^{*0}$ the following
ratios~\cite{Ali:1978kn,Cheng-Yang}:
$H_{00}:H_{--}:H_{++} \sim {\cal O}(1):{\cal O}(1/m_b):{\cal O}(1/m^2_b)$,
where $H_{00}$ corresponds to the final vector mesons in the longitudinal
polarization state, while $H_{++}, H_{--}$ in the transverse positive and
negative one. This SM result is in an obvious disagreement with the
BABAR~\cite{BABAR}, BELLE~\cite{BELLE} and CDF~\cite{CDF} observations,
demonstrating that $|H_{++}\pm H_{--}|^2\approx |H_{00}|^2$.

In the literature various efforts have been undertaken
to account for the polarization anomaly from the view point
of the SM~\cite{Cheng-Yang,Anom-SM} and in
various scenarios of new physics beyond the SM~\cite{Anom-BSM,Das:2004}.
In Ref.~\cite{Das:2004} a model independent analysis of the B-decay
polarization anomaly has been carried out on the basis of the effective
Lagrangian approach. Two sets of effective $\Delta B = 1$ operators
necessary for the resolution of this anomaly have been identified.
In addition from the experimental data~\cite{BABAR,BELLE} the
corresponding values of their Wilson coefficients have been determined.
These two sets of operators have the following chirality structure:
(i)  $(1-\gamma_5)\otimes (1-\gamma_5)$,
     $\sigma(1-\gamma_5)\otimes \sigma(1-\gamma_5)$ and
(ii) $(1+\gamma_5)\otimes (1+\gamma_5)$,
     $\sigma(1+\gamma_5)\otimes \sigma(1+\gamma_5)$.

In the present paper we use this model independent result in order
to examine the ability of R-parity violating SUSY (\rp SUSY) to
resolve the above discussed polarization anomaly in
$B^0(\bar{B}^0)\rightarrow \phi K^{*0}(\bar{K}^{*0})$-decay. In
Sec.~2 we specify the effective $\Delta B = 1$ operators satisfying
the polarization Anomaly Resolution  Criteria (pARC). In Sec.~3 we
determine these operators in the context of \rp SUSY and derive
their Wilson coefficients. In Sec.~4 we study experimental limits on
these Wilson coefficients from the existing $B_s\rightarrow
\mu^+\mu^-$ data and discuss the compatibility of the pARC with
these limits.

\section{Criteria for resolution of the polarization anomaly}

The effective Hamiltonian ${\cal H}$ describing
$\bar B^0\rightarrow \phi \bar K^{*0}$ with $\Delta B=1$
can be written in the form:
\eq\label{Eff-Hamilton}
{\cal H}_{\Delta B=1} \, = \,
{\cal H}_{\Delta B=1}^{SM} \, + \,  {\cal H}_{\Delta B=1}^{NP} \,  = \,
\frac{G_F}{\sqrt{2}}\sum_{i=1}^{14} c_i(\mu) \cdot O_i(\mu) \, + \,
 \frac{G_F}{\sqrt{2}} \sum_{i=15}^{44} c_i(\mu) \cdot O_i(\mu)
+ \ {\rm H.c.} \,,
\en
where $c_i(\mu)$ are the Wilson coefficients evaluated at the
renormalization scale $\mu\sim m_b$.
The first 14 terms correspond to the penguin-dominated SM contributions
${\cal H}_{\Delta B=1}^{SM}$ listed in Ref.~\cite{Buchalla},
the last 30 terms ${\cal H}_{\Delta B=1}^{NP}$
appear in the presence of new physics (NP).
In Ref.~\cite{Das:2004} it was shown
that out of the 30 NP-operators only the following operator set
\eq
O_{15} &=& \bar{s}_\alpha P_R b_\alpha
\cdot \bar{s}_\beta P_R s_\beta,\ \ \  \ \ \ \ \ \ \ \ \
O_{16} = \bar{s}_{\alpha}P_R b_{\beta}\cdot \bar{s}_{\beta}P_R s_{\alpha},
\label{O:15-16}\\
O_{17} &=& \bar{s}_\alpha P_L b_\alpha
\cdot \bar{s}_\beta P_L s_\beta,\ \ \  \ \ \ \ \ \ \ \ \ \,
O_{18} = \bar{s}_{\alpha}P_L b_{\beta}\cdot \bar{s}_{\beta}P_L s_{\alpha},
\label{O:17-18}\\
O_{23} &=& \bar{s}_\alpha \sigma^{\mu\nu}P_R b_\alpha
\cdot \bar{s}_\beta \sigma_{\mu\nu}P_R s_\beta,\ \ \
O_{24} = \bar{s}_{\alpha}\sigma^{\mu\nu}P_R b_{\beta}\cdot
\bar{s}_{\beta}\sigma_{\mu\nu}P_R s_{\alpha},
\label{O:23-24}\\
O_{25} &=& \bar{s}_\alpha \sigma^{\mu\nu} P_L b_\alpha
\cdot \bar{s}_\beta \sigma_{\mu\nu}P_L s_\beta,\ \ \
O_{26} = \bar{s}_{\alpha}\sigma^{\mu\nu}P_L b_{\beta}\cdot
\bar{s}_{\beta}\sigma_{\mu\nu}P_L s_{\alpha}, \label{O:25-26}
\en
satisfies the polarization Anomaly Resolution Criteria
(pARC)~\cite{Das:2004}, allowing one to possibly solve the polarization
anomaly in $B^0(\bar{B}^0)\rightarrow \phi K^{*0}(\bar{K}^{*0})$-decay.
Here, $\alpha$ and $\beta$ are the color indices.
In what follows we denote the set of operators in
Eqs.~(\ref{O:15-16})-(\ref{O:25-26})
as pARC operators. In Ref.~\cite{Das:2004} it was noted that the
(pseudo-)scalar operators $O_{15-18}$ can be expressed in the basis
of (pseudo-)tensor operators $O_{23-26}$ by Fierz transformation
\eq\label{Fierz}
O_{15}&=& \frac{1}{12}O_{23}-\frac{1}{6}O_{24}, \ \ \
O_{16}= \frac{1}{12}O_{24}-\frac{1}{6}O_{23},\\
O_{17}&=& \frac{1}{12}O_{25}-\frac{1}{6}O_{26}, \ \ \
O_{18}= \frac{1}{12}O_{26}-\frac{1}{6}O_{25}.
\en
The contributions of the operators $O_{15-26}$ to the helicity amplitudes
of $\bar B^0\rightarrow \phi \bar K^{*0}$-decay can be calculated within
the QCD factorization (QCDF) approach in terms of the corresponding Wilson
coefficients and hadronic form factors. In this approach the helicity
amplitudes take the form~\cite{Cheng-Yang,Das:2004}
\eq\label{Helicity}
\bar{H}_{00} &=& - 4 i f^T_{\phi} m^2_B(\tilde{a}_{23}
- \tilde{a}_{25})\left[h_2 T_2(m^2_{\phi})-h_3 T_3(m^2_{\phi})\right],\\
\bar{H}_{\pm\pm} &=& - 4 i f^T_{\phi} m^2_B \left\{\tilde{a}_{23}
\left[\pm f_1 T_1(m^2_{\phi})-f_2 T_2(m^2_{\phi})\right]+
\tilde{a}_{25}\left[\pm f_1 T_1(m^2_{\phi})
+ f_2 T_2(m^2_{\phi})\right]\right\} \,.
\en
Here the $\phi$-meson tensor decay constant $f^T_{\phi}$ and
the form factors of the $\bar{B}-\bar{K^*}$ transition are defined as
\eq\label{hadr-par}
\langle \phi(q,\epsilon_1)|\bar s \sigma^{\mu\nu} s |0\rangle
&=& -i f^T_{\phi} (\epsilon^{\mu *}_1 q^{\nu}-\epsilon^{\nu *}_1 q^{\mu}),\\
\langle \bar{K}^*(p',\epsilon_2)|\bar s
\sigma_{\mu\nu}q^{\nu}(1+\gamma_5) s |\bar{B}(p)\rangle &=&
2 i \epsilon_{\mu\nu\rho\sigma}\epsilon^{\nu *}_2
p^{\rho} p^{\prime \sigma} T_1(q^2)
\, + \, T_2(q^2) [\epsilon^*_{2 \mu}(m^2_B - m^2_{K^*}) \nonumber\\
&-& (\epsilon^*_2\cdot p)(p+p')_{\mu}]
+ T_3(q^2) (\epsilon^*_2\cdot p) \left[q_{\mu}
- \frac{q^2 (p+p')_{\mu}}{m^2_B - m_{K^*}^2}\right]\,,
\en
with $q = p-p^\prime$ and $m_B = 5.279$ GeV, $m_{K^*} = 0.892$ GeV,
$m_\phi = 1.019$ GeV being the masses of the $B$, $K^*$ and
$\phi$ mesons, respectively.
The kinematical factors in Eqs.~(\ref{Helicity}) are:
\eq
f_1 &=& \frac{2 p_c}{m_B},\ \ \ \ f_1=\frac{m^2_B-m^2_{K^*}}{m_B^2}\,,
\label{def-H1}\\
h_2 &=& \frac{1}{2 m_{K^*}m_{\phi}} \ \biggl[\frac{(m^2_B - m_{\phi}^2
- m_{K^*}^2)(m^2_B - m_{K^*}^2)}{m_B^2} - 4 p_c^2 \biggr] \,,
\label{def-H2}\\
h_3 &=& \frac{1}{2 m_{K^*}m_{\phi}} \
\frac{4 p_c^2 m_{\phi}^2}{m^2_B - m_{K^*}^2} \,, \label{def-H3}
\en
where
$p_c$ is the momentum of the $\phi$ or $K^*$ meson in the rest frame
of the decaying  $\bar B^0$ meson.
The effective coefficients in Eq.~(\ref{Helicity}) are expressed
in terms of the Wilson coefficients as~\cite{Das:2004}
\eq\label{a_23-def}
\tilde{a}_{23} &=& \left(1 +\frac{1}{2 N_c}\right) \left(c_{23}
+ \frac{1}{12}c_{15} - \frac{1}{6}c_{16}\right)+
\left(\frac{1}{N_c}+\frac{1}{2}\right) \left(c_{24} + \frac{1}{12}c_{16}
- \frac{1}{6}c_{15}\right) + \mbox{nonfact.}\,, \\
\label{a_25-def}
\tilde{a}_{25} &=& \left(1 +\frac{1}{2 N_c}\right) \left(c_{25}
+ \frac{1}{12}c_{17} - \frac{1}{6}c_{18}\right)+
\left(\frac{1}{N_c}+\frac{1}{2}\right) \left(c_{25} + \frac{1}{12}c_{18}
- \frac{1}{6}c_{17}\right) + \mbox{nonfact.}
\en
The last terms in~(\ref{a_23-def}) and~(\ref{a_25-def})
indicate corrections due to deviations from the QCDF.

On the basis of the above equations in Ref.~\cite{Das:2004}
there has been made an analysis of the experimental data
obtained by BABAR~\cite{BABAR} and BELLE~\cite{BELLE}
on the angular distribution in
$B^0(\bar B^0)\rightarrow \phi K^{*0}(\bar K^{*0})$-decay.
It was shown that there are two theoretical scenarios which
can separately account for the polarization anomaly of these data.\\
Scenario (i): $\tilde{a}_{23}=0$ and
\eq\label{a_25-best-fit}
|\tilde{a}_{25}| = 2.10^{+0.19}_{-0.12} \times 10^{-4}, \ \ \
\delta_{25} = 1.15\pm 0.09,\ \ \phi_{25}= -0.12\pm 0.09 \,.
\en
Scenario (ii): $\tilde{a}_{25}=0$ and
\eq\label{a_23-best-fit}
|\tilde{a}_{23}| = 1.70^{+0.11}_{-0.07} \times 10^{-4}, \ \ \
\delta_{23} = 2.36\pm 0.10,\ \ \phi_{23}= 0.14\pm 0.09 \,.
\en
Here the following notations were used $\tilde{a}_{ij}=
|\tilde{a}_{ij}|e^{i \delta_{ij}}e^{i \phi_{ij}}$, identifying
$\phi_{ij}$ and $\delta_{ij}$ with the weak (coming from the terms in
Eq.~(\ref{Eff-Hamilton})) and strong phases, respectively.
The values of Eqs.~(\ref{a_25-best-fit}) and (\ref{a_23-best-fit})
correspond to the best fit values for the combined data of
BABAR~\cite{BABAR} and BELLE~\cite{BELLE}.
In what follows we use these results as a criterion to assess if
a particular model is able to resolve the polarization anomaly in
question or not.

\section{pARC operators in \rp SUSY}

In relation to the polarization anomaly in $\bar B^0\rightarrow \phi
\bar K^{*0}$ we are studying the $\Delta B = 1$ transitions on the
quark level. Here we derive the effective Lagrangian describing
these transitions within the minimal \rp SUSY model (\rp MSSM) and
show that among the resulting set of operators there emerge the pARC
operators $O_{15}$ and $O_{17}$. In the generic case of \rp MSSM
R-parity is violated by the following terms in the superpotential
\eq
W_{\rpm} = \mu_{j}L_{j}H_{2} +
\frac{1}{2}\lambda_{ijk}L_{i}L_{j}E_{k}^{c} + \bar\lambda
_{ijk}^{\prime }L_{i}Q_{j}D_{k}^{c} + \frac{1}{2}\bar\lambda
_{ijk}^{\prime \prime }U_{i}^{c}D_{j}^{c}D_{k}^{c}, \label{sup1} \en
and in the soft SUSY breaking part of the scalar potential
\eq\label{scalar}
V_{\rpm}^{\rm soft} = \Lambda_{ijk}\tilde L_i
\tilde L_j \tilde E_k^c + \Lambda^{\prime}_{ijk}\tilde L_i \tilde
Q_j \tilde D_k^c + \Lambda^{\prime\prime}_{ijk}\tilde U_i^c \tilde
D_j^c \tilde D_k^c + \tilde\mu_{2j}^2\tilde L_j H_2 +
\tilde\mu_{1j}^2\tilde L_j H^{\dagger}_1 + \ {\rm H.c.}
\en
In Eq.~(\ref{sup1}) $L$, $Q$ stand for the lepton and quark doublet
left-handed superfields, while $E^{c},\ U^{c},\ D^{c}$ for the
lepton and {\em up}, {\em down} quark singlet superfields;  $H_{2}$
is the Higgs doublet superfields with a weak hypercharge $Y= +1$,
respectively. In Eq.~(\ref{scalar}) $\tilde{L}_i$ denotes the scalar
slepton weak doublet, $H_{1,2}$ are the scalar Higgs doublet fields.
In the above equations the trilinear terms proportional to $\lambda,
\bar\lambda^{\p}, \Lambda, \Lambda^{\p}$ and the bilinear terms
violate lepton number, while the trilinear terms proportional to
$\bar\lambda^{\p\p}, \Lambda^{\p\p}$ violate baryon number
conservation. The coupling constants $\lambda $
($\bar\lambda^{\prime\prime }$) are antisymmetric in the first
(last) two indices. The bar sign in $\bar\lambda', \bar\lambda'' $
denotes that all the definitions are given in the gauge basis for
the quark fields. Later on we will change to the mass basis and drop
the bars. Using the freedom in the definition of lepton and Higgs
superfields we choose the basis where the vacuum expectation values
of all the sneutrino fields vanish: $\langle\tilde\nu_i\rangle=0$.

The Lagrangian terms generated by the trilinear terms of the
superpotential in Eq.~(\ref{sup1}) and involving two down quarks
needed for the construction of the pARC operators
in~(\ref{O:15-16})-(\ref{O:25-26}) are as follows:
\eq\label{tril-lag}
{\cal L}_{\lambda} =  - \lambda^{\prime}_{ijk}
\tilde  \nu_{iL}\bar d_{k}P_L d_{j} - \frac{1}{2}
\lambda^{\prime\prime}_{ijk} \tilde  u^\star _{i R}\bar d_{j}P_L
d^c_{k} + \ {\rm H.c.}
\en
where $d_j$ stands for the down quark. 
It can be easily seen that the interactions in 
Eq.~(\ref{tril-lag}) can generate in second order perturbation theory 
the only $\Delta B = 1$ effective operator contributing to  
$B^0(\bar B^0)\rightarrow \phi K^{*0}(\bar K^{*0})$-decay. This is the 
operator $(\bar s P_L b)(\bar s P_R s)$ which does not belong to the 
pARC operators listed in Eqs.~(\ref{O:15-16})-(\ref{O:25-26}). Thus, 
we conclude that the trilinear \rp-couplings alone cannot resolve 
the polarization anomaly in  
$B^0(\bar B^0)\rightarrow \phi K^{*0}(\bar K^{*0})$-decay. 

Let us see if the bilinear \rp-terms may help in the solution of
this problem. The presence of the bilinear terms leads to terms in
the scalar potential which are linear in the sneutrino fields,
$\tilde{\nu}_{i}$. First, this results in $\tilde\nu-H^0_{1,2}$
mixing. Also, the linear terms drive the $\tilde{\nu}_{i}$ fields to
non-zero vacuum expectation values $\langle
\tilde{\nu}_{i}\rangle\neq 0$ at the minimum of the scalar
potential. At this ground state the MSSM vertices $\tilde{Z}\nu$ $
\tilde{\nu}$ and $\tilde{W}e\tilde{\nu}$ produce the gaugino-lepton
mixing mass terms $\tilde{Z}\nu\langle \tilde{\nu}\rangle,
 \tilde{W}e\langle \tilde{\nu}\rangle $
(with $\tilde{W},\tilde{Z}$ being wino and zino fields).
These terms taken along with the lepton-higgsino
$\mu _{i}L_{i}\tilde{H}_{1}$ mixing from the
superpotential of Eq.~(\ref{sup1}) form $7\times 7$ neutral fermion and
$5\times 5$ charged fermion mass matrices~\cite{Now}. This leads to
a non-trivial neutrino mass matrix and Lepton Flavor Violation in the
sector of charged leptons. However, these effects are obviously irrelevant
for the generation of the effective 4-quark operators.

The above mentioned effect of sneutrino-Higgs mixing $\tilde\nu-H^0_{1,2}$
is different. It corresponds to a non-diagonal mass matrix for the neutral
scalars $(H_1^0, H_2^0, \tilde\nu_e, \tilde\nu_{\mu},\tilde\nu_{\tau})$
in the bilinear part of the \rp \, scalar potential~\cite{snu-H}. 
From Eqs.~(\ref{sup1}) and (\ref{scalar}) we write: 
\eq\label{bil-scalar-pot}
V_{\rpm}^{\rm soft} = ( \mu^*\mu_j H_1^\dagger + \tilde\mu_{2j}^2  H_2
+ \tilde\mu_{1j}^2  H^{\dagger}_1 ) \tilde{L}_j + \ {\rm H.c.}
\en
Using the minimization condition 
\eq\label{mincond} 
\tilde\mu_{1j}^2 + \mu^*\mu_j + \tilde\mu_{2j}^2 \tan\beta = 0
\en
in the basis of lepton and Higgs superfields where
$\langle\tilde\nu_i\rangle=0$ we can rewrite
Eq.~(\ref{bil-scalar-pot}) in the form
\eq\label{bil-scalar-pot-1}
V_{\rpm}^{\rm soft} =
\tilde\mu_{2j}^2 (H_2 - \tan\beta H^{\dagger}_1)\tilde{L}_j
+ \ {\rm H.c.} \,,
\en
where $\tan\beta = \langle H^0_2\rangle/\langle H^0_1\rangle$.
Rotating these fields to the mass eigenstate basis we assume smallness
of sneutrino-Higgs mixing characterized by the small ratio
$(\tilde\mu_{kj}/M_{h_{1,2}})^2$, where $\tilde\mu^2_{kj}$ is the \rp
soft parameter from Eq.~(\ref{scalar}) and $M_{h_{1,2}}$ are the neutral
Higgs masses~\cite{Gunion}. In the leading order in this small parameter
we obtain the following interactions of sneutrinos with down quarks and
charged leptons
\eq\label{snu-h}
{\cal L}_{\tilde\nu ll} = \eta_{j} \
\left[\frac{m_{d_i}}{M_W}(\bar{d}_i\ d_i)
+ \frac{m_{l_i}}{M_W}(\bar{l}_i\ l_i)\right]\tilde\nu_j \,,
\en
with the couplings
\eq\label{SH-par}
\eta_{j} =  \frac{g_2}{2} \tilde\mu_{2j}^2
\frac{\tan\beta}{\sqrt{1+\tan^2\beta}}
\left(\frac{\cos\alpha}{M_{h_2}^2}-\frac{\sin\alpha}{M_{h_1}^2}\right) \,.
\en
Here $\alpha$ is the mixing angle of the neutral Higgses in the limit of
no mixing with the sneutrino fields:
\eq\label{Higgs-mix}
H_1^0 = -\sin\alpha\cdot h_1^0 + \cos\alpha\cdot h_2^0,\ \ \
H_2^0 = \cos\alpha\cdot h_1^0 + \sin\alpha\cdot h_2^0 \,,
\en
where $h^0_{1,2}$ are the corresponding mass eigenstates with the masses
$M_{h_1}, M_{h_2}$.
Note that $H^0_2$, which has no couplings to the down quarks and leptons,
does not contribute to Eq.~(\ref{snu-h}).

Now, combining the trilinear and bilinear \rp-interactions from
Eq.~(\ref{tril-lag}) and Eq.~(\ref{snu-h}), as shown in Fig.1, we obtain
in second order perturbation theory the following effective Hamiltonian
after integrating out the heavy sneutrino fields:
\eq\label{b-dll}
{\cal H}_{\rpm} &=&
\frac{m_{d_j}}{M_W} (\bar{d}_j\ d_j)\left(\frac{\eta_i}{m_{\tilde\nu_i}^2}
\lambda^{\p *}_{im3}\ \bar{d}_m P_R b + \frac{\eta^*_i}{m_{\tilde\nu_i}^2}
\lambda^{\p}_{i3m}\ \bar{d}_m P_L b\right) + \nonumber\\
&+& \frac{m_{l_j}}{M_W} (\bar{l}_j\ l_j)
\left(\frac{\eta_i}{m_{\tilde\nu_i}^2} \lambda^{\p *}_{im3} \
\bar{d}_m P_R b + \frac{\eta^*_i}{m_{\tilde\nu_i}^2}
\lambda^{\p}_{i3m} \ \bar{d}_m P_L b\right) + \ {\rm H.c.}
\en
The 4-quark terms involve the pARC operators $O_{15}$ and $O_{17}$ from
the list of Eqs.~(\ref{O:15-16})-(\ref{O:25-26})
with the following Wilson coefficients:
\eq\label{c_15_17}
c_{15}= \frac{\sqrt{2}}{G_F}\frac{m_{s}}{M_W}
\frac{\eta_i}{m_{\tilde\nu_i}^2} \lambda^{\p *}_{i23} \,, \ \ \
c_{17}= \frac{\sqrt{2}}{G_F}\frac{m_{s}}{M_W}
\frac{\eta^*_i}{m_{\tilde\nu_i}^2} \lambda^{\p}_{i32} \,.
\en

Thus \rp SUSY seems to satisfy the pARC as it allows
appropriate operator structures.
In the following we have to check if the existing
experimental constraints on the \rp-parameters entering into
the definition of the Wilson coefficients allow one to accommodate
the values of Eqs.~(\ref{a_25-best-fit}) and (\ref{a_23-best-fit}).

\section{Experimental Constraints on Wilson coefficients}

Examining Eq.~(\ref{b-dll}) we note that the strength of both
the 4-quark and quark-lepton operators is determined by the same
combination of the R-parity conserving and \rp-parameters forming the
Wilson coefficients $c_{15,17}$. Therefore, one can directly constrain
the $c_{15,17}$ parameters from the existing stringent experimental upper
bound on the $B_s\rightarrow \mu^+\mu^-$ branching ratio~\cite{CDF-Bmumu}
\eq\label{Lim1}
Br(B_s\rightarrow \mu^+\mu^-)\leq 1.0\times 10^{-7} \
( 90\% \ {\rm C.L.} ) \,.
\en
An important advantage of this constraint is that it applies to the
coefficients $c_{15,17}$ as a whole, avoiding uncertainties related
to the presence of several R-parity conserving
$(\tan\beta, \alpha, M_{h_1,h_2}, m_{\tilde\nu})$ and
violating parameters $(\tilde\mu_{2j}^2, \lambda')$.

The contribution of the quark-lepton interactions in the
Lagrangian (\ref{b-dll}) to the decay rate of this process
can be written in terms of the Wilson coefficients $c_{15,17}$ as 
\eq\label{Br}
\Gamma(B_s\rightarrow \mu^+\mu^-)= \frac{G_F^2}{2}\frac{m_{B_s}}{32 \pi}
\left(\frac{m_{\mu}}{m_{s}}\right)^2
\left(f_{B_s}\frac{m^2_{B_s}}{m_b+m_s}\right)^2 (c_{15}-c_{17})^2
\left[1-\left(\frac{2 m_{\mu}}{m_{B_s}}\right)^2\right]^{3/2}
\en
where we used
\eq\label{hme1}
\langle 0|\bar{s} \gamma_5 b|\bar B_s^0\rangle =
 i f_{B_s} \frac{m^2_{B_s}}{m_b+m_s} \,.
\en
We use the following numerical values for the quantities in above
equations: $f_{B_s} = 0.2$~GeV \cite{f_bs}, $m_{B_s}=5.367$~GeV,
$m_b =4.6$~GeV, $m_s = 0.15$~GeV and
$\tau_{B_s}= 1.46\times 10^{-12} $s~\cite{PDG}.
Considering the two scenarios of Ref.~\cite{Das:2004}
as displayed in Eqs.~(\ref{a_25-best-fit}) and (\ref{a_23-best-fit})
[denoted as (i) and (ii)] we get from the experimental
limit~(\ref{Lim1}) the following upper bounds
\eq\label{lim-c}
|c_{15}|, |c_{17}| \leq 1.4\times 10^{-4}.
\en
Using the definitions of Eqs.~(\ref{a_23-def}) and (\ref{a_25-def})
these limits can be translated to upper limits on the effective
coefficients
\eq\label{lim-a}
|\tilde{a}_{23}|, |\tilde{a}_{25}| \leq 5.9\times 10^{-6}.
\en
These limits are about two orders of magnitude smaller then
the values given in Eqs. (\ref{a_25-best-fit}) and (\ref{a_23-best-fit})
required for the solution of the polarization anomaly.

Thus, we conclude that the polarization anomaly observed in
$B^0(\bar B^0)\rightarrow \phi K^{*0}(\bar K^{*0})$ decay
by the BABAR~\cite{BABAR} and BELLE~\cite{BELLE} collaborations
cannot be explained within the \rp SUSY framework, despite
the occurrence of effective operators with the chiral structure
required qualitatively.

As a byproduct of our analysis the limits of Eq.~(\ref{lim-c})
set new upper limits on the products of the soft and superpotential
\rp-parameters of Eqs.~(\ref{SH-par}) and (\ref{c_15_17}).
Since the expressions for the Wilson coefficients $c_{15,17}$ contain
the R-parity conserving parameters as well we choose one representative
point in the SUSY parameter space in order to illustrate the limits on
the \rp-parameters. We take a typical mSUGRA:
the so-called SPS 1a point from the list of nine Snowmass benchmark
points \cite{SPS}. This choice corresponds to
$\tan\beta = 10$, $m_0=-A_0=0.25 m_{1/2}=100$ GeV and $\mu>0$.
For this parameters we find
\eq\label{RPV-limit}
\left(\frac{\tilde\mu_{2 i}}{100 \ \mbox{GeV}}\right)^2
|\lambda^{\prime}_{i23}|, \
\left(\frac{\tilde\mu_{2 i}}{100 \ \mbox{GeV}}\right)^2
|\lambda^{\prime}_{i32}|
\leq 5.6 \times 10^{-3}.
\en

To our knowledge in the literature (for a review see, for 
instance~\cite{RPV-rev}) there have not been established
experimental limits on these products of \rp-parameters. 
However, there exist bounds on $\tilde\mu_{2i}^2$,
$\lambda^{\prime}_{i23}$ and $\lambda^{\prime}_{i32}$ separately 
from various low energy processes ~\cite{RPV-rev}. This allows one to
obtain indirect bounds on their products and compare them with those 
in Eq.~(\ref{RPV-limit}). The soft \rp-parameter $\tilde\mu_{2i}^2$,
contributes to the neutrino mass matrix at one-loop level. 
Thus it is constrained by the present limits on neutrino masses and 
mixing from neutrino oscillations. With the SPS 1a set of the R-parity 
conserving parameters one has: 
$(\tilde\mu_{2i}/100 \ {\rm GeV})^2\leq 10^{-4}$.  
Existing constraints on the trilinear \rp-couplings 
are typically as follows: 
$\lambda^{\prime}_{i23}, \lambda^{\prime}_{i32} \leq 0.2$. 
Combining these constraints we have the limits:
\eq\label{RPV-limit-lit}
\left(\frac{\tilde\mu_{2 i}}{100 \ \mbox{GeV}}\right)^2
|\lambda^{\prime}_{i23}|, \
\left(\frac{\tilde\mu_{2 i}}{100 \ \mbox{GeV}}\right)^2
|\lambda^{\prime}_{i32}|
\leq 2.0 \times 10^{-5}.
\en
which are two orders of magnitude better then those in 
Eq.~(\ref{RPV-limit}). Nevertheless, the latter can still 
be useful as direct constraints on the specific products of 
the bilinear and trilinear \rp-parameters. 
Note that these constraints correspond to a particular point 
in the MSSM parameter space and in some other points the above 
limits may significantly change. The detailed study of this 
question is beyond the scope of the present paper.

\section{Conclusions}

We analyzed the \rp SUSY model with respect to its ability to
account for the polarization anomaly in $B^0(\bar B^0)\rightarrow
\phi K^{*0}(\bar K^{*0})$-decay observed by the BABAR \cite{BABAR}
and BELLE \cite{BELLE} collaborations. Within this framework we have
determined the effective $\Delta B=1$ operators with chirality
structures appropriate for a possible resolution of this anomaly.
However, the experimental data on $B\rightarrow \mu^+\mu^-$-decay
set stringent limits on the respective Wilson coefficients, which
are about two orders of magnitude below the values required to
resolve the polarization anomaly. This gap of two orders of
magnitude can hardly by eliminated by the uncertainties in the
hadronic parameters involved in the calculation of the helicity
amplitudes of $B^0(\bar B^0)\rightarrow \phi K^{*0}(\bar
K^{*0})$-decay. Therefore, we do not believe that \rp SUSY is able
to account for the B-decay polarization anomaly.

As a byproduct we used the experimental data on
$B\rightarrow \mu^+\mu^-$-decay to set a new upper limit on the product
of the two \rp-parameters
$\tilde\mu^2_{2i} \, |\lambda^{\prime}_{i23}|$ and
$\tilde\mu^2_{2i} \, |\lambda^{\prime}_{i32}|$,
where $\tilde\mu^2_{2i}$ and $\lambda^{\prime}_{ijk}$ are bilinear
soft and trilinear superpotential \rp-parameters, respectively.

\vspace*{1cm}

{\bf Acknowledgments}\\[3mm]
\hspace*{1cm}
This work was supported in part by CONICYT (Chile) under grants
FONDECYT No.1030244 and PBCT/No.285/2006 and
by the DFG under contracts FA67/31-1 and GRK683.
This research is also a part of the EU Integrated
Infrastructure Initiative Hadronphysics project under the contract
No. RII3-CT-2004-506078 and President grant of Russia
"Scientific Schools"  No. 5103.2006.2.
SK would like to thank Deutscher Akademischer Austausch Dienst (DAAD)
for the support within the program ``F\" orderung ausl\" andischer
Gastdozenten zu Lehrt\" atigkeiten an deutschen Hochschulen".
SK is also thankful to the T\"ubingen theory group for hospitality.
J.C.H. thanks High Energy Physics LatinAmerican-European Network (HELEN)
for the support within the Advanced Training program.

\begin{figure}
\begin{center}
\vspace*{1cm} \epsfig{file=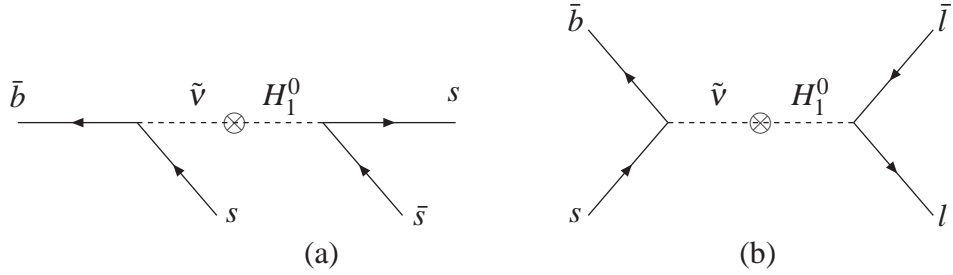, scale=1.0}
\end{center}

\caption{The \rp SUSY contribution to the $\bar{b}\rightarrow ss\bar{s}$
(a) and to the $\bar{b} s \rightarrow l \bar{l}$ transition operators.
The sign $\otimes$ denotes \rp soft sneutrino-Higgs mixing.
The left hand vertices in both diagrams are due to
the \rp superpotential $\lambda^{\prime}$ coupling,
while the right hand ones correspond to the R-parity conserving
$H_1-q-\bar{q}$ and $H_1-l-\bar{l}$ Yukawa couplings.}
\end{figure}

\end{document}